\documentclass[prd,floatfix,twocolumn]{revtex4}
\pdfoutput=1
\usepackage{hyperref}

\usepackage{natbib,ifthen}

\usepackage{amssymb} 
\usepackage{amsmath}
\usepackage{graphicx}
\usepackage{subfigure}

\usepackage{color}
\def\be{\begin{equation}}
\def\ee{\end{equation}}
\def\bea{\begin{eqnarray}}
\def\eea{\end{eqnarray}}

\def\lsim{\mathrel{\mathstrut\smash{\ooalign{\raise2.5pt\hbox{$<$}\cr\lower2.5pt\hbox{$\sim$}}}}}
\def\gsim{\mathrel{\mathstrut\smash{\ooalign{\raise2.5pt\hbox{$>$}\cr\lower2.5pt\hbox{$\sim$}}}}}

\def\({\left(}
\def\){\right)}

\def\bwt{\begin{widetext}}
\def\ewt{\end{widetext}}

\begin{document}
\bibliographystyle{prsty}
\title{Enhanced Peculiar Velocities in Brane-Induced Gravity}	
\author{Mark Wyman$^{1}$}
\author{Justin Khoury$^{2}$}

 \affiliation{$^{1}$ Perimeter Institute for Theoretical Physics, Waterloo, Ontario N2L 2Y5, Canada\\ 
 $^{2}$ Center for Particle Cosmology, University of Pennsylvania, Philadelphia, PA 19104, USA}
\date{\today}
\begin{abstract}
The mounting evidence for anomalously large peculiar velocities in our Universe presents a challenge for the $\Lambda$CDM paradigm.
The recent estimates of the large scale bulk flow by Watkins {\it et al.} are inconsistent at the nearly $3\sigma$ level with $\Lambda$CDM predictions. Meanwhile,  Lee and Komatsu have recently estimated that  the occurrence of high-velocity merging systems such as the Bullet Cluster (1E0657-57) is unlikely at a  $6.5 - 5.8\sigma$ level, with an estimated probability between $3.3\times 10^{-11}$ and $3.6\times 10^{-9}$ in $\Lambda$CDM cosmology.
We show that these anomalies are alleviated in a broad class of infrared-modifed gravity theories, called brane-induced gravity, in which gravity becomes higher-dimensional at ultra large distances. These theories include additional scalar forces that enhance gravitational attraction and therefore speed up structure formation at late times and on sufficiently large scales. The peculiar velocities are enhanced by $24-34$\% compared to standard gravity, with the maximal enhancement nearly consistent at the $2\sigma$ level with bulk flow observations. The occurrence of the Bullet Cluster in these theories
is $\approx 10^4$ times more probable than in $\Lambda$CDM cosmology.
\end{abstract}
\maketitle

A striking discrepancy with $\Lambda$CDM predictions has emerged in recent estimates of the local large scale bulk flow.
Using a compilation of peculiar velocity surveys, Watkins {\it et al.}~\cite{hudson1,hudson2} finds a bulk flow of $407\pm 81$~km~s$^{-1}$ on 50~$h^{-1}$Mpc scales, which is inconsistent at the $\approx 3\sigma$ level with the $\Lambda$CDM rms expectation of $\sim$180~km/s.

Evidence for this anomaly on larger scales, though less reliable, comes from measurements of cluster peculiar velocities using the kinetic Sunyaev-Zel'dovich effect. These indicate a coherent bulk motion of 600-1000~km~s$^{-1}$ out to $\gsim 300h^{-1}$~Mpc~\cite{kash1,kash2}. These results match the direction of the flow found by \cite{hudson1} on the length scales where they overlap. See also~\cite{lavaux} for recent efforts in reconstructing the local group peculiar velocity. In the future, observations beyond our local area will offer better statistics on bulk flows~\cite{Song:2010vh}.

Independent evidence that structure is evolving more rapidly than expected is the ``Bullet Cluster" 1E0657-57~\cite{barrena}.
A key input in hydrodynamical simulations of this system~\cite{bulletsim1,bulletsim2,bulletsim3} is the initial subcluster
velocity. Recent simulations have shown that an initial velocity of 3000~km/s at $5$~Mpc separation is required to
explain the data~\cite{mastro}. Using horizon scale N-body simulations, Lee and Komatsu~\cite{Lee:2010hj} have estimated that the probability of having such large velocities in $\Lambda$CDM cosmology is between $3.3\times 10^{-11}$ and $3.6\times 10^{-9}$ --- that is, the ``Bullet" system is between 6.5 and 5.8$\sigma$ away from the mean velocity for colliding clusters. A previous estimate of this likelihood has been interpreted as evidence for a new attractive force in the dark sector~\cite{glennysDM}. See~\cite{markevitch,bradac} for examples of other violent merging systems.

These anomalies motivate us to study peculiar velocities in a broad class of infrared (IR)
modified gravity theories called brane-induced gravity~\cite{DGP,DGP2,DGP3,DGP4,DGP5,cascade1,cascade2,cascade3,ghostfree,aux1,aux2}.
See~\cite{wetterich} for a useful parametrization of modified peculiar velocities, and~\cite{alessandra} for related work.
The theories of interest involve extra scalar degrees of freedom in $4D$, which are inherited from the higher-dimensional massless graviton.
These degrees of freedom couple to the trace of the matter stress-energy tensor and therefore enhance the effective gravitational attraction compared to Newtonian gravity at late times and on large scales. Large scale structure is more developed and is presently evolving faster on large scales than in the $\Lambda$CDM model, both of which lead to larger
bulk flows~\cite{niayeshghazal,markw}. 

While we focus on higher-dimensional theories for concreteness, enhanced peculiar velocities are also expected in any theory
with a long range extra scalar force, such as galileon
scalar-tensor theories~\cite{galnic,galvik,galileon,deser,galileoncosmo1,galileoncosmo2,galileoncosmo3}, interacting
dark sector models~\cite{DMDE0,DMDE1,DMDE2,huey, pierste,DMDE3,ravi} and symmetron theories~\cite{symmetron,symmetron2}.
In chameleon/$f(R)$~\cite{f(R),f(R)2,f(R)3,cham1,cham2,cham3,cham4} models, however, the scalar fifth force has a range $\lsim$~Mpc and
hence cannot explain the large scale anomalies discussed here. It would be interesting to study whether our results also apply to the recent IR-modified theory
proposed by~\cite{fedomodif}.

\subsection{Brane-Induced Gravity}

The principal motivation for modifying General Relativity (GR) at ultra-large distances is the cosmological constant problem~\cite{weinberg}.
(See~\cite{bhuvreview} for a recent review of cosmological tests of gravity.) Vacuum energy is the zero-momentum component of stress energy and hence its backreaction depends sensitively on the nature of gravity in the far infrared. A compelling approach is {\it degravitation}, in which gravity acts as a high-pass filter~\cite{dilute,ADGG,degrav}: the cosmological term is in fact large,
in accordance with field theory expectations, but gravitates very weakly.

A perennial challenge in devising consistent IR-modified theories of gravity is quantum stability, {\it i.e.} avoidance of ghost-like (negative energy) instabilities.
Giving the graviton a hard mass \`a la Pauli-Fiertz~\cite{FP}, for instance, unavoidably leads to instabilities~\cite{BoulwareDeser,nima,Creminellipaper}. 
A more promising approach is brane-induced gravity~\cite{DGP, DGP2,DGP3,DGP4,DGP5,cascade1,cascade2,claudiareview,nonFP,intersecting,cascade3,ghostfree,gigashif,aux1,aux2,nishant}, which relies on branes and extra dimensions. The most widely known example is the Dvali-Gabadadze-Porrati (DGP) brane-world model~\cite{DGP,DGP2,DGP3}.
The normal branch of the DGP model is perturbatively ghost free, in contrast to the self-accelerating branch~\cite{luty,nicolis,dgpghost}, and thus represents a perturbatively consistent IR modification of gravity.

The Cascading Gravity framework~\cite{cascade1,cascade2,intersecting,cascade3,ghostfree} extends the DGP model to $D\geq 6$ space-time dimensions. 
In the simplest version with a $6D$ bulk space-time, our 3-brane is embedded in a 4-brane, each with their own induced gravity terms.
The upshot of this generalization is twofold. First, the soft mass term for the graviton is a more slowly varying function of momentum than in DGP, which
is a necessary condition for degravitation~\cite{degrav}. Thus Cascading Gravity is a promising framework for
realizing this phenomenon~\cite{cascade3}. Furthermore, the cascading graviton mass term results in an expansion history that closely resembles
$\Lambda$CDM cosmology and is therefore less constrained by observations~\cite{claudiaandrew,niayeshghazal}. 

What about perturbative stability? Perturbing around $6D$ Minkowski space with empty branes reveals a ghost scalar mode.
Early work~\cite{cascade1} revealed, however, that the ghost is excised by including a sufficiently large tension
on the 3-brane or, alternatively, by considering a higher-dimensional Einstein-Hilbert term localized on the brane~\cite{cascade2,gigashif,massimo1,massimo2}.
While the original derivation of~\cite{cascade1} was restricted to a particular decoupling limit of the theory, recently the absence of perturbative ghosts has been proven rigorously
by perturbing the full $6D$ solution in the presence of brane tension~\cite{ghostfree}. These results establish the cascading gravity
framework as a perturbatively consistent IR modification of gravity.

\subsection{Summary of Results}

In this paper we show that the observational bulk flow anomaly is alleviated in brane-induced gravity theories.
The bulk flow enhancement depends on the number $D$ of bulk space-time dimensions and
the cross-over scale $r_c$ beyond which gravity on the brane becomes higher-dimensional. 
Our bulk flow results can be summarized in a fitting formula for the rms of the 1-dimensional velocity on $50h^{-1}$~Mpc scales,
which is valid for $D \geq 4$ and $r_c\lsim 1.5 H_0^{-1}$:
\be
\label{vfitintro}
v_{1-{\rm dim}}^{\rm G}\simeq 102 \times \frac{3 \gamma}{2} \( \frac{9}{5} r_c H_0 \)^{-\left(\gamma-\frac{2}{3}\right)}{\rm km/s}\,,
\ee
where 
\be
\gamma \equiv \frac{1}{6}\left(-1+\sqrt{\frac{49D-146}{D-2}}\right)\,.
\label{gam}
\ee
The superscript ``G" indicates the use of a Gaussian window function.
In particular, for $D=4$ ({\it i.e.}, $\gamma = 2/3$), this matches the $\Lambda$CDM prediction: $v_{1-{\rm dim}}^{\rm G} = 102$~km/s for our fiducial parameter choices.
In Sec.~\ref{de} we generalize~(\ref{vfitintro}) to a fitting formula valid on a range of scales --- see~(\ref{vfitintro2}). Our fiducial cosmology is consistent
with WMAP 7-year data~\cite{Larson:2010gs} and assumes
a spatially-flat universe with $\Omega_{\rm m} = 0.24$, $\Omega_{\rm b}=0.042$, a primordial power spectrum
with tilt $n_s = 0.96$, and a primordial amplitude chosen to yield a present-day normalization of  $\sigma_8 = 0.8$ for a $\Lambda$CDM growth history.

The bulk flow enhancement~(\ref{vfitintro}) grows with increasing $D$, since more extra dimensions imply more scalar fields on the brane contributing to the
gravitational attraction. It also grows with decreasing $r_c$, since smaller $r_c$ implies that departures from standard $4D$ gravity turn on
at earlier times. Because bulk flows measure today's evolution, higher-$D$ models generate bigger departures from
 $\Lambda$CDM for fixed late-time normalization, since the force enhancement grows with $D$. 

\begin{figure}[htbp] 
   \centering
   \includegraphics[width=0.5\textwidth]{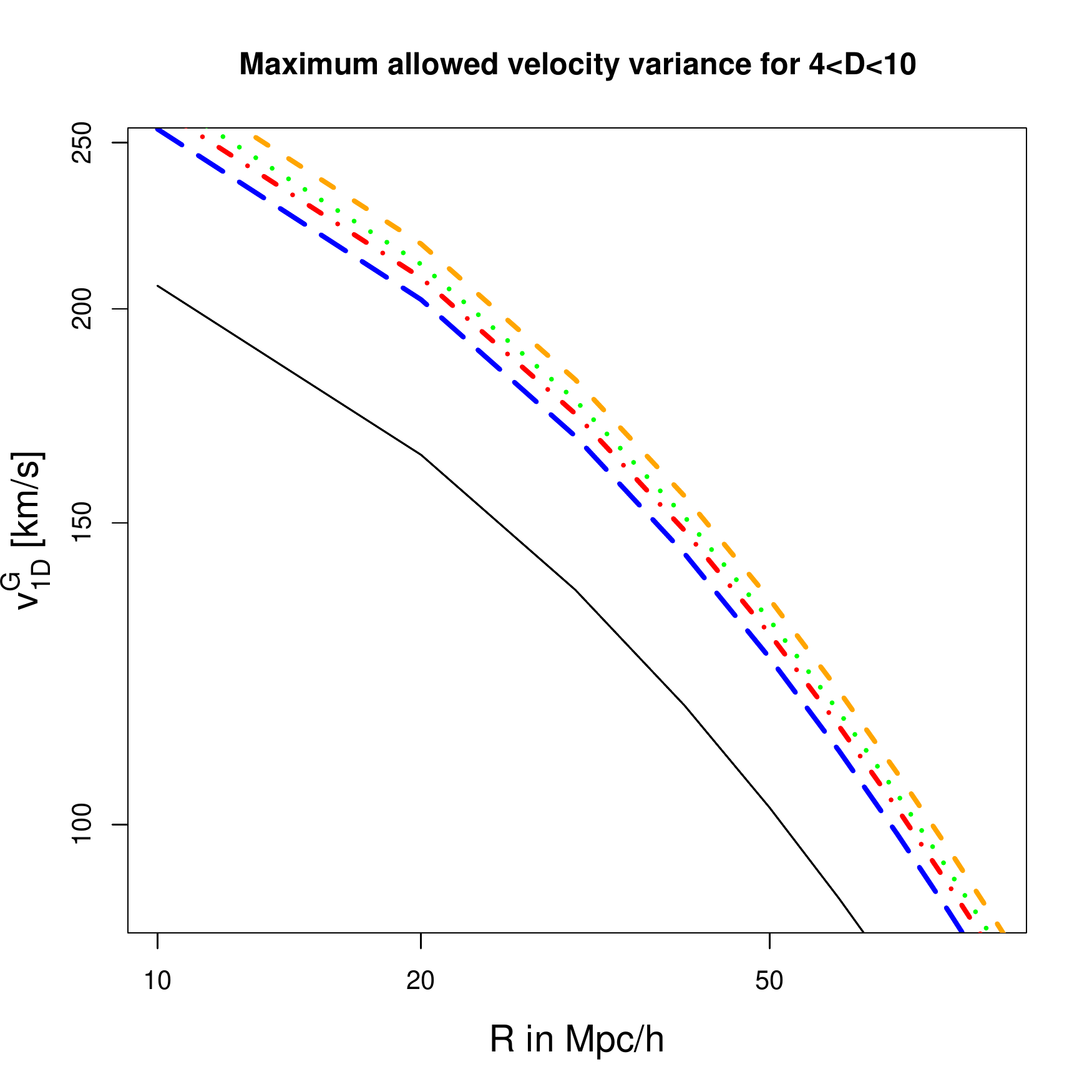} 
   \caption{The 1-dimensional rms velocity as a function of scale for standard gravity (black line) and brane-induced gravity, derived from linear analysis.
   The dashed (blue),
   dash-dotted (red), dotted (green), and top dashed (orange) lines show the maximum
   velocity allowed in our model with $D=5$, 6, 7, and 10, 
   respectively, for late-time power spectra with $\sigma_8 \simeq 0.88$, consistent with observational constraints.
    The expansion history and primordial normalization are identical in all cases.}
   \label{maxima}
\end{figure}

The enhanced gravitational attraction in cascading gravity also boosts the amplitude of density perturbations, resulting in
a larger $\sigma_8$ at late times versus standard gravity for fixed primordial normalization. The
late time $\sigma_8$ is constrained by various observables, such as cluster abundance~\cite{Vikhlinin:2008ym, Lueker:2009rx},
the weak lensing power spectrum~\cite{Fu:2007qq} and galaxy clustering~\cite{urosbias}. We discuss these constraints in Sec.~\ref{ampcons}.
The tightest constraint comes from X-ray galaxy cluster counts from the ROSAT All-Sky Survey~\cite{Vikhlinin:2008ym}, which for $\Lambda$CDM growth history imposes
$\sigma_8 \lsim 0.88$ at the 95\% C.L. 
We therefore require that the late-time amplitude in our models approximately satisfies this bound. For a given $D$, this gives a lower bound on the allowed $r_c$. On one hand, this is conservative: because our additional scalar force turns off in high-density environments, like galaxy clusters, the amplitude of the late-time {\it linear} power spectrum, when compared with data using $\Lambda$CDM methods, overestimates the cluster abundance~\cite{marcos}. On the other hand, the additional scalar force, if not completely screened, may increase the dynamical mass of any given cluster; this could increase the expected number of clusters at fixed dynamical mass, thereby tightening constraints. See~\cite{Schmidt:2010jr} for a careful study of dynamical effects in a variety of models. A similar treatment in our class of theories is work in progress~\cite{marcos}.

Figure~\ref{maxima} shows the 1-dimensional rms velocity $v_{1-{\rm dim}}^{\rm G}$ (the variance from zero mean of the theoretical distribution) as a function of scale for standard gravity and brane-induced/Cascading Gravity with $D = 5,6,7$ and 10, derived from linear theory using a Gaussian window function. For each $D$, we choose the minimum allowed value of $r_c$, so each of the modified gravity curves have a similar late-time $\sigma_8$.

To compare with data from the peculiar velocity surveys analyzed by Watkins {\it et al.}~\cite{hudson1}, we use a window function that includes observational effects but is only valid on $50h^{-1}$~Mpc scales.
Velocities obtained with this window function are denoted with superscript ``W".
In practice, this results in a $\sim 1$\% increase compared to the Gaussian window function, through the inclusion of some higher-$k$ modes. Over the range $5 \leq D < \infty$, we find
\be
220 < v_{3-{\rm dim}}^{\rm W} < 237~{\rm km/s} \,.
\label{vrange}
\ee
This is compared with the $\Lambda$CDM value, $179$~km/s. The upper end of this range is nearly consistent
at the $2\sigma$ level with the observed bulk flow of $407 \pm 81$~km/s~\cite{hudson1}. 
While we have used linear theory to derive these results, we also perform N-body simulations of the non-linear 
evolution; these are in excellent agreement with the linear analysis for the scales of interest ({\it cf.} \S \ref{numerics}). 

The stronger effective gravitational attraction in our models also makes the occurrence of a high-velocity merging system like the Bullet Cluster
much more probable. Assuming that the majority of the infall velocity is caused by the $\simeq 10^{15} M_\odot$ main cluster, we treat the clusters as point particles released from rest from a large initial separation ($30$~Mpc) and calculate
the resulting velocity at $5$~Mpc, the initial separation for gas collision simulations. 
To include the rest of large scale structure, we add a bulk flow component to the infall velocity in quadrature.

We find that the resulting velocity at $5$~Mpc separation is 14\% ($D=5$) to 27\% ($D=10$) larger in our theories,
where we again take the minimum allowed $r_c$ for each $D$. For $D = 10$ (and $r_c = 2750$~Mpc),
an initial velocity of $3000$ km/s at $z=0$ is a 4.8$\sigma$ event, as compared with 6.5$\sigma$ in $\Lambda$CDM; or, in terms of probability,
$6.6 \times10^{-7}$ versus $3.3\times 10^{-11}$, a boost of over $10^4$.
Meanwhile, at $z=0.5$, closer to the actual merger redshift of
$z=0.296$, the modified gravity result is 3.9$\sigma$ from the mean, versus 5.8$\sigma$ in standard gravity.
The probability of achieving that velocity is thus increased to $5.1\times10^{-5}$, compared to $3.6\times 10^{-9}$, again a $\sim10^4$ probability enhancement.
 If the required initial velocity is reduced
to 2000~km/s, the probability of such an occurrence in our model becomes $\sim 0.8$\% at $z=0$ and 14.2\% at $z=0.5$, respectively
257 and 65 times more likely than the $\Lambda$CDM expectation (that is, respectively 2.4 and 1$\sigma$ in modified gravity, versus 4 and 2.9$\sigma$ 
in standard gravity).

\subsection{Modeling Cascading Cosmology}

Extracting exact cosmological predictions from a higher-dimensional set-up, such as Cascading Gravity, is
technically very challenging; even the precise form of the modified Friedmann equation is not yet known.
Hence our results must rely on some phenomenological input.

Fortunately, the scales of interest for this study are well within the Newtonian regime. In this regime, the theory admits a local description
on the 3-brane, which arises through a certain decoupling limit. In this limit, the complexities of the full
higher dimensional theory that are irrelevant to $4D$ physics are left out. The result is
an effective theory in $4D$ with new degrees of freedom. The relevant degrees of freedom are weak-field gravity and
$D-4$ scalar fields coupled to the trace of the matter stress-tensor, describing brane-bending along each extra dimension.
Whereas gravity is weakly coupled in this limit, the scalars have non-linear derivative interactions that are responsible for
the Vainshtein effect, a phenomenon generic to this kind of theory that leads to the approximate recovery of standard gravity in high density regions. While the precise form of the
decoupling theory for Cascading Gravity is not known, we draw upon known results in DGP to infer 
the non-linear interactions of the scalars. 

For the background, meanwhile, we assume a $\Lambda$CDM expansion history. As mentioned earlier, the form of the modified graviton propagator
in Cascading Gravity suggests that brane-world corrections to the Friedmann equation~\cite{cedricbrane,cline,branereview,renjie} are more slowly-varying functions of $Hr_c$ than in standard DGP~\cite{cedric}; they should closely resemble vacuum energy contributions. Furthermore, by assuming a $\Lambda$CDM expansion history, our analysis isolates the effects of the modified growth history~\cite{bhuvpenjie}. 
We describe this approach in detail in Sec.~\ref{massgrav}.

Finally, we wish to emphasize that, although our analysis does not constitute a fully rigorous derivation of cosmological
predictions in Cascading Gravity, our results should most likely capture the essence of the predictions of the complete
theory. Because the observables of interest are in the Newtonian regime, the description in terms of conformally-coupled
scalar fields with non-linear interactions should be accurate. The details of the Vainshtein screening mechanism may vary,
but the corrections to our predictions --- which lie on scales where linear theory is valid ---  are expected to be small. In short, our approach is not merely a toy model of cascading cosmology,
but a first iteration in extracting predictions from a broad class of IR-modified gravity theories.

\section{Massive/Resonance Gravity}
\label{massgrav}

The defining feature of theories with infinite-volume extra dimensions, such as DGP and Cascading Gravity, is that 4$D$ gravity is mediated by a resonance graviton --- a continuum of massive states --- with general propagator
\be
\frac{1}{k^2 +m^2(k)}\,.
\label{prop}
\ee
In DGP, for instance, $m^2(k) = r_c^{-1}k$.  In real space, this gives a gravitational potential that interpolates from
the $4D$ scaling, $1/r$, at short distances to the $5D$ scaling, $1/r^2$, at large distances, with a cross-over scale set by $r_c$.
In Cascading Gravity, the soft mass term $m^2(k)$ is a more complicated function of $k$, involving multiple
cross-over scales~\cite{cascade1,cascade2}. For simplicity, we shall assume that all cross-over scales are comparable and denote this common scale by $r_c$. 

Because $4D$ gravity is massive, each graviton has 5 polarizations: the usual 2 helicity-2 states of GR, 2 helicity-1 states and 1 helicity-0 degree of freedom.
At distances $r\ll r_c$, only the helicity-2 and helicity-0 degrees of freedom are relevant --- the helicity-1 states are very weakly coupled to matter and can be safely ignored.
In the DGP model, the helicity-0 mode has a nice geometrical interpretation. It measures the extrinsic curvature of the brane in the extra dimension and is thus
referred to as a brane-bending mode.
Cascading Gravity theories have $D-5$ additional helicity-0 or scalar modes, accounting for the higher number of extra dimensions in which the brane can bend. This counting of degrees of freedom can alternatively be understood from a decomposition of the $D$-dimensional massless spin-2 representation~\cite{cascade1,nonFP}.

These scalar modes couple to the trace of the stress-energy tensor of matter on the brane and, combined with the helicity-2 states, result in
a one-graviton exchange amplitude between conserved sources having the tensor structure of $D$-dimensional massless
gravity~\cite{cascade2}:
\be
{\cal A} \sim T^{\mu\nu} \cdot \frac{1}{k^2}\cdot \tilde{T}_{\mu\nu} - \frac{1}{D-2}T^\mu_{\; \mu} \cdot \frac{1}{k^2}\cdot \tilde{T}^\nu_{\; \nu}\,,
\label{amp}
\ee
where we have neglected the graviton mass term for the scales of interest~\cite{footnote1}. 
For non-relativistic sources, this corresponds to the modified Poisson equation
\be
k^2\Psi_{\rm dyn} = -4\pi G \left(1 + \frac{D-4}{D-2}\right) \rho \,.
\label{poissonflat}
\ee
In other words, the gravitational attraction is a factor of $1+\frac{D-4}{D-2}$ stronger than in standard Newtonian gravity.

If~(\ref{amp}) were valid in the solar system, the theory would already be ruled out by post-Newtonian constraints, for
arbitrarily small $m$ --- this is the famous van Dam-Veltman-Zhakarov (VDVZ) discontinuity~\cite{vDVZ}. The 
resolution, first proposed by Vainshtein for massive gravity~\cite{vainshtein}, is that the weak-field/linear assumption
implicit in~(\ref{amp}) is actually a poor approximation for the scalar modes in the vicinity of
astrophysical bodies or in the early universe. Instead, as shown explicitly in the DGP model~\cite{resolve,resolveothers}, non-linearities in these modes in high-density regions
result in them decoupling from matter, leading to an approximate recovery of Einstein gravity in, {\it e.g.}, the solar system. 

Let us start with the DGP case. On scales $\ll H^{-1}$ (and thus $\ll r_c$), we can neglect time derivatives relative to gradients of the graviton helicity-0 mode.
This mode, denoted by $\chi$, satisfies the approximate equation~\cite{lue,Fabian,Roman}
\bea
\nonumber
& & \nabla^2 \chi + \frac{r_c^2}{3\beta_{\rm DGP}a^2} [ (\nabla^2\chi)^2
- (\nabla_i\nabla_j\chi)(\nabla^i\nabla^j\chi) ] \\
&& \qquad \qquad\qquad\qquad= \frac{8\pi Ga^2\delta\rho}{3\beta_{\rm DGP}} \,,
\label{pieqn}
\eea
where the $\nabla$'s denote spatial derivatives, and where
\be
\beta_{\rm DGP} \equiv 1 + 2 Hr_c\left(1+\frac{\dot{H}}{3H^2}\right)\,.
\label{betaDGP}
\ee
The overdot denotes a derivative with respect to proper time.
The Vainshtein effect appears in two guises in the above. Firstly, in the early universe when $Hr_c \gg 1$, 
the coupling to matter density becomes vanishingly small since $\beta_{\rm DGP} \gg 1$.
Secondly, even at late times when $Hr_c\lsim 1$ and $\beta_{\rm DGP} \approx1$,
sufficiently large overdensities trigger non-linearities in $\chi$ and result in its decoupling.

The analogue of~(\ref{pieqn}) is not yet known for Cascading Gravity. There are multiple scalars in this case, each 
of which is expected to exhibit Vainshtein screening. There have not yet been
any successful calculations that keep the non-linearities of all scalar modes. For the purpose of this paper, we shall take a
phenomenological approach and assume that all cascading scalar degrees of freedom obey an equation of
the form~(\ref{pieqn}). This is consistent with our earlier assumption of a single cross-over scale $r_c$.
In other words, we collectively denote the scalars by $\chi$ and assume that they satisfy
\bea
\nonumber
& & \nabla^2 \chi + \frac{r_c^2}{3\beta a^2} [ (\nabla^2\chi)^2
- (\nabla_i\nabla_j\chi)(\nabla^i\nabla^j\chi) ] \\
&& \qquad \qquad\qquad\qquad= \frac{8\pi G a^2\delta\rho}{3\beta}\,.
\label{pieqn2}
\eea
Using the scaling of the resulting deviation from GR in the solar system~\cite{degrav}, we can infer
that the cosmological deviation scales as $\beta \sim H^2r_c^2$. We then generalize~(\ref{betaDGP}) to
\be
\beta \equiv 1 + 2 H^2r_c^2\left(1+\frac{\dot{H}}{3H^2}\right)\,.
\label{beta}
\ee
This modification makes a sharper transition in time than~(\ref{betaDGP}). This is related to the fact that gravity rapidly
weakens on long length scales in models that can successfully degravitate the cosmological constant. This also
means that whatever replaces~(\ref{pieqn2}) in Cascading Gravity will likely exhibit even more
efficient Vainshtein screening within collapsed objects.
The effect of the $\chi$ field on matter is through its contribution to the actual potential $\Psi_{\rm dyn}$ that moves particles:
\be
\Psi_{\rm dyn} = \Psi_{\rm N} + \frac{3}{2} \(\frac{D-4}{D-2} \)\chi\,,
\label{dyn}
\ee
where $\Psi_{\rm N}$ is the usual Newtonian potential. The $D$-dependence here is chosen to recover~(\ref{poissonflat}) in the linearized limit. When performing N-body simulations, we solve~(\ref{pieqn2}) using a multi-grid relaxation scheme similar to the one
described in~\cite{Fabian}. Our N-body code is an updated version of the one used in previous work~\cite{markw}, revised to
solve~(\ref{pieqn2}) exactly and without resorting to the approximation of spherical symmetry. Meanwhile, we solve for $\Psi_{\rm N}$ in the usual way
through Fourier transforms on a particle mesh. The code and further numerical results will be described in more detail elsewhere~\cite{marcos}.

Most of our results are derived within linear theory, since this is a valid approximation for bulk flows over the scales of interest.
In this regime, we can obtain a modified evolution equation for the density perturbations $\delta\equiv \delta\rho/\bar{\rho}$. Using the fact
that the energy-momentum tensor on the brane is covariantly conserved, density and velocity perturbations evolve as usual via
\be
\dot{\delta} = -\frac{k}{a}v\;;\qquad \dot{v} + Hv = \frac{k}{a}\Psi_{\rm dyn}\,.
\label{emomcons}
\ee
These can be combined with the linearized version of~(\ref{pieqn}) to yield
\be
\ddot{\delta} + 2H\dot{\delta} = 4\pi G \bar{\rho} \left(1+\beta^{-1}\frac{D-4}{D-2}\right)\delta\,.
\label{deltaevolve}
\ee
Note that this is consistent with~(\ref{poissonflat}) in the flat space limit $\beta\rightarrow 1$. As another quick check,
letting $\beta\rightarrow\beta_{\rm DGP}$ and setting $D=5$, this agrees with explicit cosmological results in DGP~\cite{lue}.
In this paper, we will solve for peculiar velocities using linear theory and cross-check the results with simulations.

\section{Review of bulk flow formalism}
\label{bulkrev}

Since matter responds simply to gravitational gradients at the linear level, it is straightforward
to estimate the expected bulk flows on length scales where matter overdensities
are in the linear regime. See \S II.14 of~\cite{Peebles} for an introductory exposition. 

The root-mean-square of the one-dimensional (1-dim) velocity, $\sigma_v^2 \equiv \langle v_{\rm 1-dim}^2\rangle$,
on a scale $R$ is given by~\cite{Watkins:1997tq} 
\be
\label{sigv}
\sigma_v^2(R) = \frac{1}{3} \cdot \frac{H_0^2 }{2 \pi^2} \int_0^\infty P(k) W^2(k,R)f^2(k) {\rm d}k\,,
\ee
where $P(k)$ is the power spectrum of matter density fluctuations, and $W(k)$ is a window function with scale $R$
(To get the three-dimensional answer, we simply remove the $1/3$).
Similarly to~\cite{Watkins:1997tq}, we choose the window function to be a Gaussian for the majority of our analysis:
\be
W^{\rm G}(k) = \exp(-k^2 R^2)\,.
\label{gauss}
\ee
This is designed to capture only the small-$k$/long-distance behavior of the power spectrum, {\it i.e.} the bulk flow. 
Note that the window function used for determining the bulk flow from peculiar velocity surveys \cite{hudson1} is slightly different from this, including some higher-$k$/smaller-$R$
modes. We present this observational window function in Sec.~\ref{data} and use it when comparing with the
results of~\cite{hudson1}. In practice, however, it gives very similar results to the Gaussian window function.
See Fig.~\ref{wind} for a sneak preview of these window functions.

The velocity also depends on the growth rate $f(k)$ of perturbations, 
\be
f(k) \equiv \left. \frac{{\rm d}\ln g(a,k)}{{\rm d}\ln a}\right\vert_{a = a_0}\,,
\label{growthrate}
\ee
where $g$ is the growth function, and $a_0$ is the present scale factor. In the gravity theories of interest (as well as in theories with clustering dark energy)
both $g$ and $f$ are generically scale-dependent.

This formalism is similar to that used to set the normalization of the matter power spectrum. 
The parameter most commonly used for this purpose is $\sigma_8$, 
the matter fluctuation within $8 h^{-1}$ Mpc spheres:
\be
\sigma_8^2 \equiv \frac{1}{2\pi^2} \int_0^\infty k^2 \, P(k) W^2_8(k) {\rm d}k\,,
\ee
where $W_8(k) = 3j_1(kR_8)/kR_8$, with $R_8 = 8 h^{-1}$ Mpc and $j_1$ a spherical Bessel function. Note that
the window function for $\sigma_8$ is a shallower function of $k$ than that for $\sigma_v$.
Hence, $\sigma_8$ folds in more higher-$k$ modes, whereas peculiar velocities are a direct probe of
low-$k$ power.

The long-distance modifications of gravity of interest boost peculiar velocities in two ways:
\begin{itemize}
\item Faster development of structure at late times, encoded in the growth rate $f$;
\item Greater build-up of structure due to the integrated influence 
of stronger gravity, resulting in a larger amplitude of the power spectrum ({\it i.e.}, larger $\sigma_8$). 
\end{itemize}
As mentioned earlier, however, the amplitude of the late-time matter power spectrum cannot be too drastically altered,
as it is constrained by large-scale structure observations. These will be taken into account in Sec.~\ref{ampcons} to constrain our model parameters. 
Nevertheless, the boost in the bulk flow can be substantial because of the growth rate effect. In other words, whereas the power spectrum is the integrated
result of the entire growth history, peculiar velocities are also sensitive to the present growth rate.

Bulk flows are often discussed in a variety of different ways, which can be confusing.
The bulk flows that are measured using peculiar velocity surveys and 
other techniques are reconstructions of a full three-dimensional bulk flow of our region of space,
but they are based on measurements that are separately only one-dimensional ---
our line-of-sight measurements always project the velocity of each object onto the radial direction.
What is measured, then, is a 1-dimensional velocity for each object in the survey.
By collecting and averaging over a large number of objects, however, we are able to reconstruct
the full 3D bulk flow~\cite{hudson1}, albeit with a small residual contribution
from some smaller scales. This can be seen in the difference between our window function~(\ref{gauss}) and the ideal experimental window
function for surveys~(\ref{winfit}), shown in Fig. \ref{wind}, which includes a ``bump" at around $k\sim 0.05h$~Mpc$^{-1}$.

Present observational techniques only allow reliable measurement of our local bulk flow. 
Thus, we have three independent measurements to compare with the expected variance
in the local velocity calculated in~(\ref{sigv}). After reviewing the results of our determination
of the variance in Secs.~\ref{est}-\ref{numerics}, we compare those theoretical expectations
with the current best local bulk flow data in Sec.~\ref{data}.
Although it is an effect on a different length scale, we also discuss how our model
change expectations for Bullet cluster initial velocities in light 
of~\cite{Lee:2010hj} in Sec.~\ref{data}.

\section{Analytical Estimates} \label{est}

In this Section, we estimate the expected bulk flows in Cascading Gravity, first through analytical methods assuming matter-dominated cosmology (Sec.~\ref{md}) and
through fitting formulas of linear theory integration for $\Lambda$CDM expansion history (Sec.~\ref{de}). 

\subsection{Enhancement in Matter-Dominated Cosmology}
\label{md}

In the linear regime and on scales much smaller than $r_c$, density perturbations evolve according to~(\ref{deltaevolve}).
To get a rough estimate of the expected bulk flow enhancement analytically, we first ignore dark energy and
consider an Einstein-de Sitter ($\Omega_{\rm m} =1$) universe. 

At early times ($Hr_c \gg 1$), the cosmological Vainshtein effect results in $\beta \gg 1$, and density perturbations evolve as in standard gravity,
with growing mode solution $\delta\sim a$. Once $Hr_c \lsim 1$, however, the extra scalar modes of Cascading Gravity become effective and $\beta \approx 1$.
Equation~(\ref{deltaevolve}) then reduces to
\be
\ddot{\delta} + \frac{4}{3t}\dot{\delta} -\frac{4}{3t^2}\left(1+\frac{D-4}{D-2}\right)\delta = 0\,,
\ee
with growing mode solution
\be
\delta \sim t^\gamma\,,
\label{delgam}
\ee
where $\gamma$ is given by~(\ref{gam}). As a quick check, since $\gamma = 2/3$ for $D=4$, this reproduces the usual matter-dominated
growth $\delta \sim a$. 

Since $H\sim t^{-1}$ in matter-dominated cosmology, the excess growth from the onset of the
modified gravity phase (when $H = 1/r_c$) until the present time is
\be
\delta  = \delta^{\rm std\;grav} \times (H_0r_c)^{-\gamma+\frac{2}{3}}\,.
\label{deltaenhance}
\ee
Through the continuity equation,~(\ref{delgam}) translates into $v\sim \dot{\delta} = \gamma \delta/t$ for the
peculiar velocity. The enhancement relative to standard gravity is thus
\be
v = v^{\rm std\;grav} \times \frac{3\gamma}{2} \cdot (H_0r_c)^{-\gamma+\frac{2}{3}}\,.
\label{vpecenhance}
\ee
This expression neatly captures the two contributions to the bulk flow excess described in Sec.~\ref{bulkrev}: the larger growth rate, through the $3\gamma/2$ factor,
and the boost in the power spectrum amplitude through~(\ref{deltaenhance}). The above derivation, which crudely assumes a sharp onset of the modified growth when
$H = r_c^{-1}$, actually agrees to within a few \% with the exact integration of~(\ref{deltaevolve}) for matter-dominated cosmology. This illustrates the rapid turn-on of the scalar force
for Cascading Gravity,~(\ref{beta}).

\subsection{Corrections from including dark energy}
\label{de}

The above analytic estimate, while useful in guiding our thinking, does not capture the weakening of growth
triggered by the onset of cosmic acceleration. Using a $\Lambda$CDM expansion history in~(\ref{deltaevolve}) 
is straightforward but requires numerical integration. The resulting growth rate, $f(a,k)$, and
power spectrum, $P(k)$, are then substituted back into~(\ref{sigv}) to obtain the velocity dispersion $\sigma_v$
as a function of scale $R$. In comparing modified and standard gravity, we keep other
cosmological parameters fixed at their fiducial values: $\Omega_{\rm m}=0.24$, $\Omega_{\rm b}=0.042$, $h=0.73$, $n_s=0.96$, 
and a primordial amplitude chosen to yield $\sigma_8=0.80$ for $\Lambda$CDM growth history, consistent with WMAP 7-year data.

In a similar spirit to Peebles'  famous expression for the growth rate in the presence of dark energy,
$f_{\Lambda {\rm CDM}} \simeq \Omega_{\rm m}^{\;0.6}$, we use the results of numerically solving the linear perturbation equations to derive a fitting formula for
the effect of our modifications to gravity on the amplitude of clustering today. For $D = 5,\ldots 10$ and over the range $500 < r_c < 7000$~Mpc, we find that a reasonable fit is given by
\be
\label{fit}
\sigma_8  \simeq \sigma_8^{\rm std\;grav} \times \left(\frac{2}{3}H_0 r_c\right)^{0.71\left(-\gamma+\frac{2}{3}\right)}\,.
\ee
We can also use a fitting technique to extend the expression for the peculiar velocity on $50h^{-1}$~Mpc scales given by~(\ref{vfitintro})
to a formula valid over a range of scales $R \ll r_c$:
\bea
\label{vfitintro2}
&v_{1-{\rm dim}}^{\rm G}(R) \simeq  102\times \frac{3 \gamma}{2} \( \frac{9}{5} r_c H_0 \)^{-\left(\gamma-\frac{2}{3}\right)} \times \nonumber \\
 &\quad  \quad e^{-23 \left ( \( \frac{50h^{-1} {\rm Mpc}}{R\;[h^{-1}{\rm Mpc}]}\)^{0.1}\hspace{-0.05in}- 1\right )}\hspace{-0.05in} \( \frac{50h^{-1} {\rm Mpc}}{R\;[h^{-1}{\rm Mpc}]}\)^{4/5} \hspace{-0.12in}{\rm km/s} \,,
\eea
with $\gamma$ defined in~(\ref{gam}). 

\section{Constraints on the Amplitude}
\label{ampcons}

The boost in the amplitude of density perturbations translates into a larger value for $\sigma_8$ compared
to standard gravity, for fixed primordial amplitude. It is important to emphasize that the precise, early-time WMAP 
limit on $\sigma_8$ applies only to the primordial amplitude as evolved to today using standard growth history;
our choice of this amplitude, $\sigma_8 =0.8$, is consistent with WMAP 7-year results \cite{Larson:2010gs}. 
The amplitude of the late-time power spectrum is measured by various observations discussed below,  constraining our models.

\begin{figure}[htbp] 
   \centering
   \includegraphics[width=0.5 \textwidth]{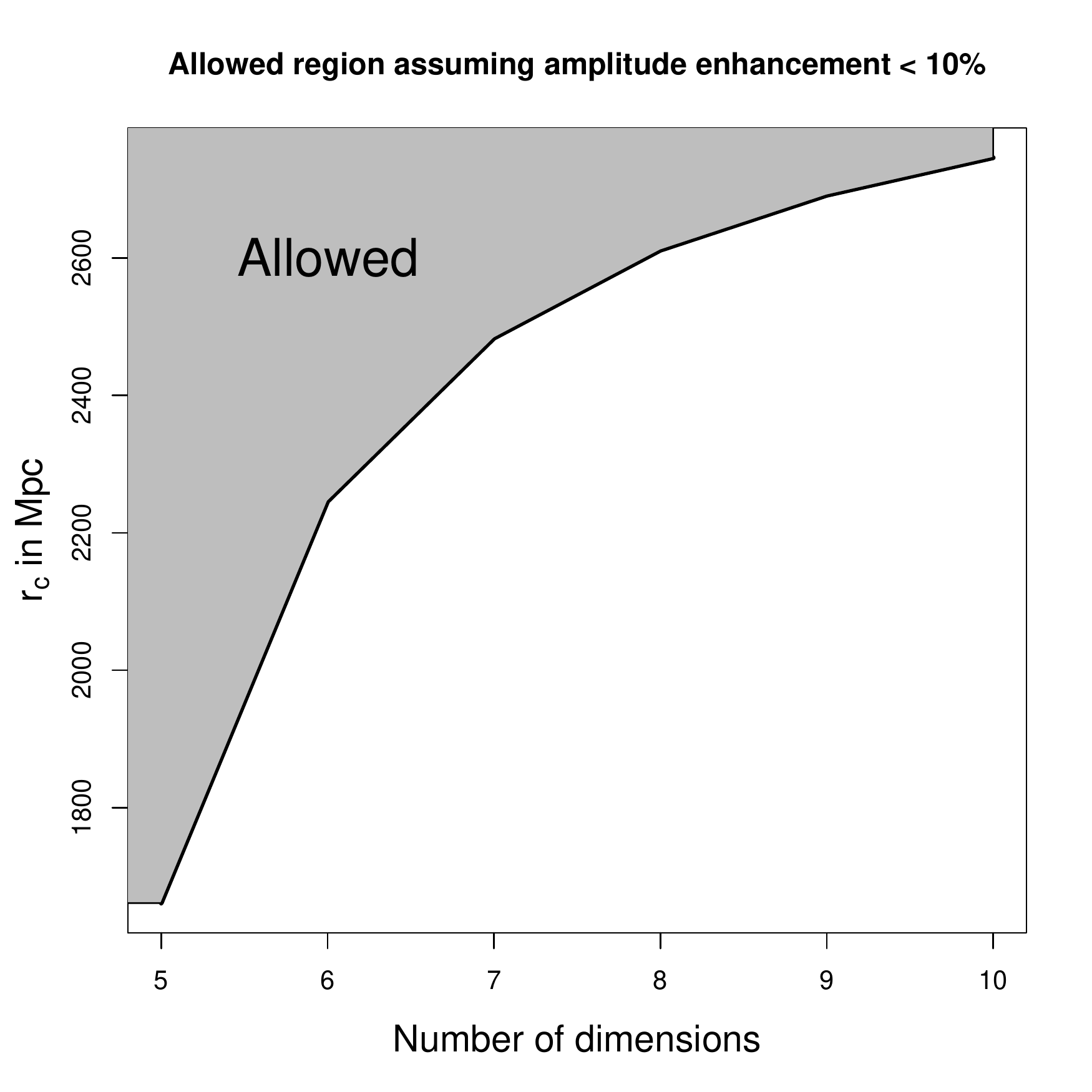} 
   \caption{The shaded region of the $D-r_c$ parameter space shows the values allowed
   by observational constraints on the power spectrum amplitude. See also Table~\ref{tab1}.}
      \label{fudge}
\end{figure}

Notable omissions in our discussion are large-scale tests, such as the cosmic microwave background temperature anisotropy and
the Integrated Sachs-Wolfe galaxy cross-correlation. In DGP, these observables are the most constraining for $r_c$~\cite{lamdgpobs}.
However, analysis of these effects requires evolving perturbations on horizon scales; this is beyond our Newtonian treatment. Though our decoupling-limit arguments (as in Sec. \ref{massgrav}) should give robust predictions on Newtonian scales, horizon-scale observables require a relativistic treatment of cosmological perturbation theory in Cascading Gravity. 

\begin{table}[htbp]
   \centering
   \begin{tabular}{|c|cc|}
   \hline
 \multicolumn{3}{|c|}{     $D$  \; $r_c$ (in Mpc)} \\ \hline
     5      && 1665  \\ \hline
     6         &  & 2250      \\ \hline
      7      & & 2486  \\ \hline
      8       && 2614 \\ \hline
      9 & &2694  \\ \hline
      10    &   & 2750 \\
      \hline
   \end{tabular}
   \caption{Minimum values for $r_c$ in Mpc allowed by observational constraints on the power spectrum amplitude.
   This contains the same information as Fig.~\ref{fudge}. }
   \label{tab1}
\end{table}

\subsection{X-ray Clusters}

The tightest constraint comes from cluster counts using X-ray observations of the ROSAT All-Sky Survey~\cite{Vikhlinin:2008ym}:
\be
\sigma_8 \( \frac{\Omega_{\rm m}}{0.24} \)^{0.47} = 0.829 \pm 0.0275 \quad \mbox{(Clusters)}\,,
\label{clusters}
\ee
where the error bar includes a 9\% systematic uncertainty in the mass calibration~\cite{Vikhlinin:2008ym}. For our fiducial $\Omega_{\rm m} = 0.24$,
this implies $\sigma_8 \lsim 0.88$ at the 95\% C.L. The translation from cluster abundance observations to $\sigma_8$ assumes
standard gravity, while in our theories the dependence of the mass function on the linear power spectrum amplitude is modified. We will come back to
this point shortly. In this work we take the $\sigma_8$ constraint at face value and, with our fiducial choice $\sigma_8^{\rm std\;grav} = 0.8$, 
impose $\sigma_8/\sigma_8^{\rm std\;grav} \lsim 1.1$. Using~(\ref{fit}), this translates into a constraint on $r_c$ and $D$:
\be
\left(\frac{2}{3}H_0 r_c\right)^{0.71\left(-\gamma+\frac{2}{3}\right)}  \lsim 1.1\,.
\label{sig8bound}
\ee
The allowed region of the $D-r_c$ parameter space is shown in Fig.~\ref{fudge}. Note that for a given number of space-time
dimensions $D$, this translates into a lower bound on $r_c$. In particular, we find $r_c \gsim 1665$ and 2750~Mpc for $D=5$
and 10, respectively. 

The above constraint is, on the one hand, conservative. Because our scalar force turns off in regions of high density, such as galaxy clusters, we expect fewer and smaller
non-linear structures in our model as compared with a $\Lambda$CDM model with identical present-day power spectrum normalization. On the other hand, the additional
scalar force in our model leads to systematically higher dynamical masses in clusters \cite{Schmidt:2010jr}. This implies an increase in the number of clusters of fixed
dynamical mass as compared with $\Lambda$CDM with the same late-time $\sigma_8$. The bound we use~(\ref{sig8bound}) assumes that these opposing
effects approximately cancel. These considerations have been studied in detail by~\cite{Schmidt:2009am} for chameleon/$f(R)$ cosmology. A similar treatment in Cascading Gravity
is work in progress~\cite{marcos}.

Substituting the {\it minimum} allowed value of $r_c$ for each $D$ in~(\ref{vfitintro2}), we obtain the maximum allowed bulk flow
as a function of scale. The result is shown in Fig.~\ref{maxima} for $D = 5,6,7$ and 10. On scales $50h^{-1}$~Mpc probed
by~\cite{hudson1},  using the appropriate observational window function (see Sec.~\ref{data}), we find the range~(\ref{vrange}):
\be
220 < v_{3-{\rm dim}}^{\rm W} < 237~{\rm km/s} \,.
\label{vrange2}
\ee
This should be contrasted with the $\Lambda$CDM prediction of $179$~km/s. The upper range of~(\ref{vrange2}) is therefore almost consistent
at the $2\sigma$ level with the observed bulk flow of $407 \pm 81$~km/s~\cite{hudson1}. In Sec.~\ref{data} we present a more careful comparison with the data.

\subsection{Weak Lensing}

The Canadian France Hawaii Telescope weak Lensing Survey (CFHTLS)~\cite{Fu:2007qq} finds 
\be
\sigma_8 \( \frac{\Omega_{\rm m}}{0.24} \)^{0.53} = 0.855 \pm 0.086 \quad \mbox{(Weak Lensing)}\,.
\label{WLcons}
\ee 
This is a much weaker constraint than~(\ref{clusters}) and is automatically satisfied for the range of late-time
values for $\sigma_8$ considered here. As a side remark, note that the constraint on our modified gravity theories
from weak lensing is actually weaker than implied by~(\ref{WLcons}) because of the cosmological Vainshtein effect.
The lensing kernel for CFHT sources spans the redshift range $0.25 \leq z \leq 1$, where the effect of modifications to
gravity is somewhat muted: for $D=10$ and $r_c = 2750$ Mpc, the case that yields the largest deviation from standard
gravity, the difference from standard gravity varies from $\lsim 5$\% at $z=1$ to $\approx$7.5\% at $z=0.5$. 
Hence the average enhancement of $\sigma_8$ for observations made in this redshift range is 
less than the $z=0$ result suggests.

\subsection{Galaxy Clustering}

The amplitude of the late time matter power spectrum is also constrained by the clustering of galaxies measured by redshift surveys, such as the Sloan Digital Sky Survey (SDSS)~\cite{SDSS} and the 2 Degree Field Galaxy Redshift Survey (2dFGRS)~\cite{2dF}. However, an immediate comparison of these observations is not straightforward due to the uncertainty in the galaxy bias. Over the mass range of interest, though, preliminary results using N-body simulations show little difference in halo bias in our model compared with the standard gravity prediction~\cite{marcos}. Marginalizing over bias, Seljak {\it et al.}~\cite{urosbias} found 
\be
\sigma_8 = 0.88 \pm 0.06  \quad \mbox{(Galaxies)}\,.
\label{galax}
\ee
Again this is a much less restrictive result than the X-ray cluster observations. 
It is worth noting that the value of $\sigma_8$ that we use in our modified
gravity results, which is in considerably better agreement with the peculiar velocity
data than $\Lambda$CDM, coincides with the central value of~(\ref{galax}).

\subsection{$E_g$ Parameter}

The expectation value of the ratio of galaxy-galaxy lensing to galaxy-velocity cross-correlations has been proposed as an observational test of gravity~\cite{rachelEg}. In linear theory, this combination, denoted by $E_g$, is independent of bias and initial power spectrum normalization. 
Using SDSS luminous red galaxies at $z=0.32$,~\cite{urosEg} recently obtained
\be
E_g = 0.392 \pm 0.065\,.
\label{Egobs}
\ee
In both $\Lambda$CDM cosmology and Cascading Gravity, this parameter is given by
\be
E_g = \frac{\Omega_{\rm m}}{f}\,,
\ee
where $\Omega_{\rm m}$ is the present-day matter density, and $f = {\rm d}\ln g/{\rm d}\ln a$ is the growth rate at the redshift of observation.
With our fiducial value of $\Omega_{\rm m} = 0.24$, the $\Lambda$CDM result is $E_g = 0.387$. Our IR-modified gravity theories predict
$0.333 \lsim  E_g \lsim 0.346$ for $10\geq D \geq 5$, assuming as before the minimum allowed $r_c$ for each $D$. Values in this range are consistent with~(\ref{Egobs}).

\begin{figure}[htbp] 
   \centering
   \includegraphics[width=0.5 \textwidth]{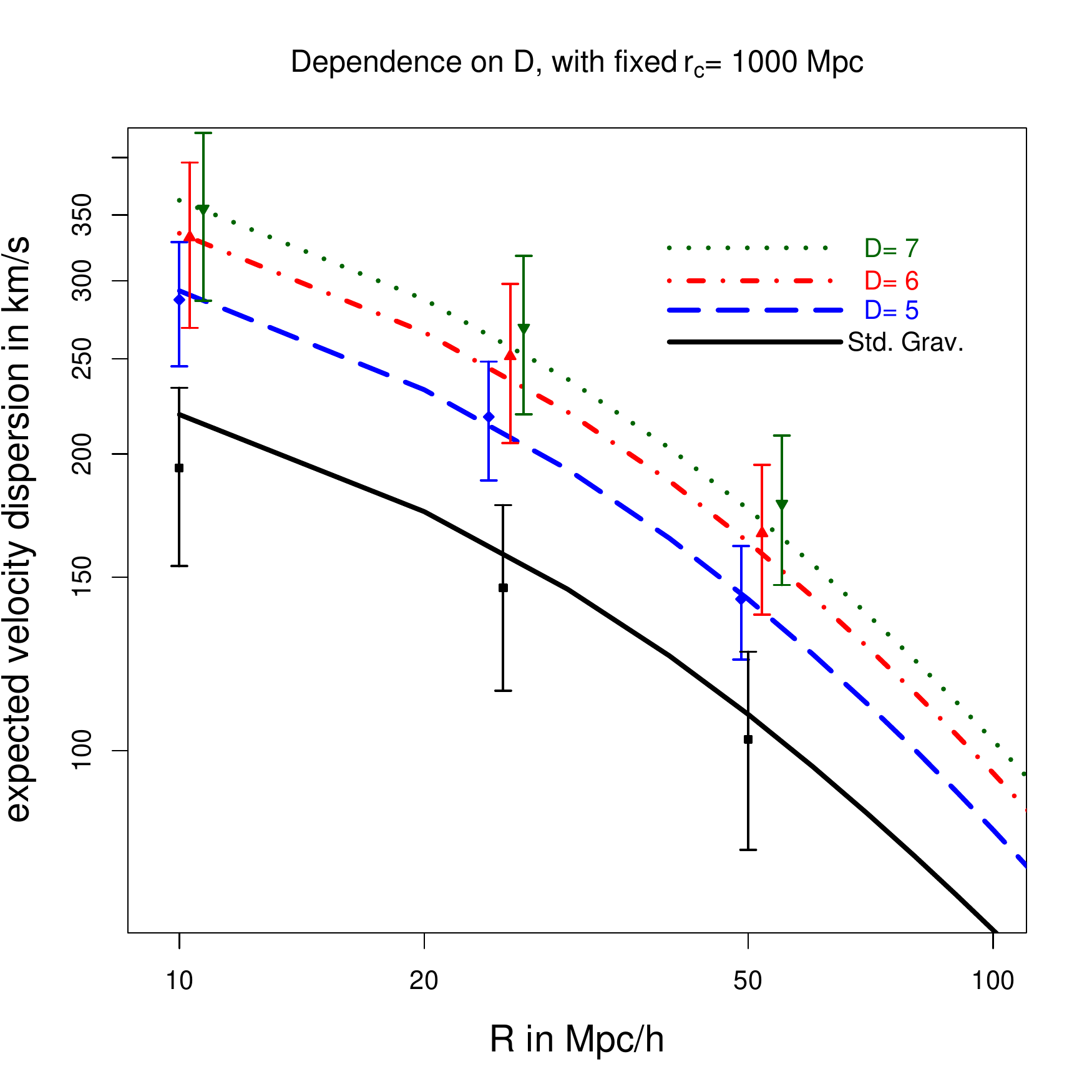} 
   \caption{The expected 1-dimensional peculiar velocity as a function of scale, using the Gaussian window function~(\ref{gauss}). The curves are from linear theory whereas points (with bootstrap-estimated error bars)
   are from N-body simulations. The black curve/points are the standard gravity results. The long dashed (blue), dash-dotted (red) and dotted (green) curves/points are respectively the $D=5,6$ and $7$ results, keeping $r_c =1000$~Mpc fixed. The points at a given scale are slightly offset from one another for readability. For this comparison, $\sigma_8[\Lambda\mbox{CDM}]=0.85$.}
 \label{peclin1000}
\end{figure}

\begin{figure}[htbp] 
   \centering
   \includegraphics[width=0.5 \textwidth]{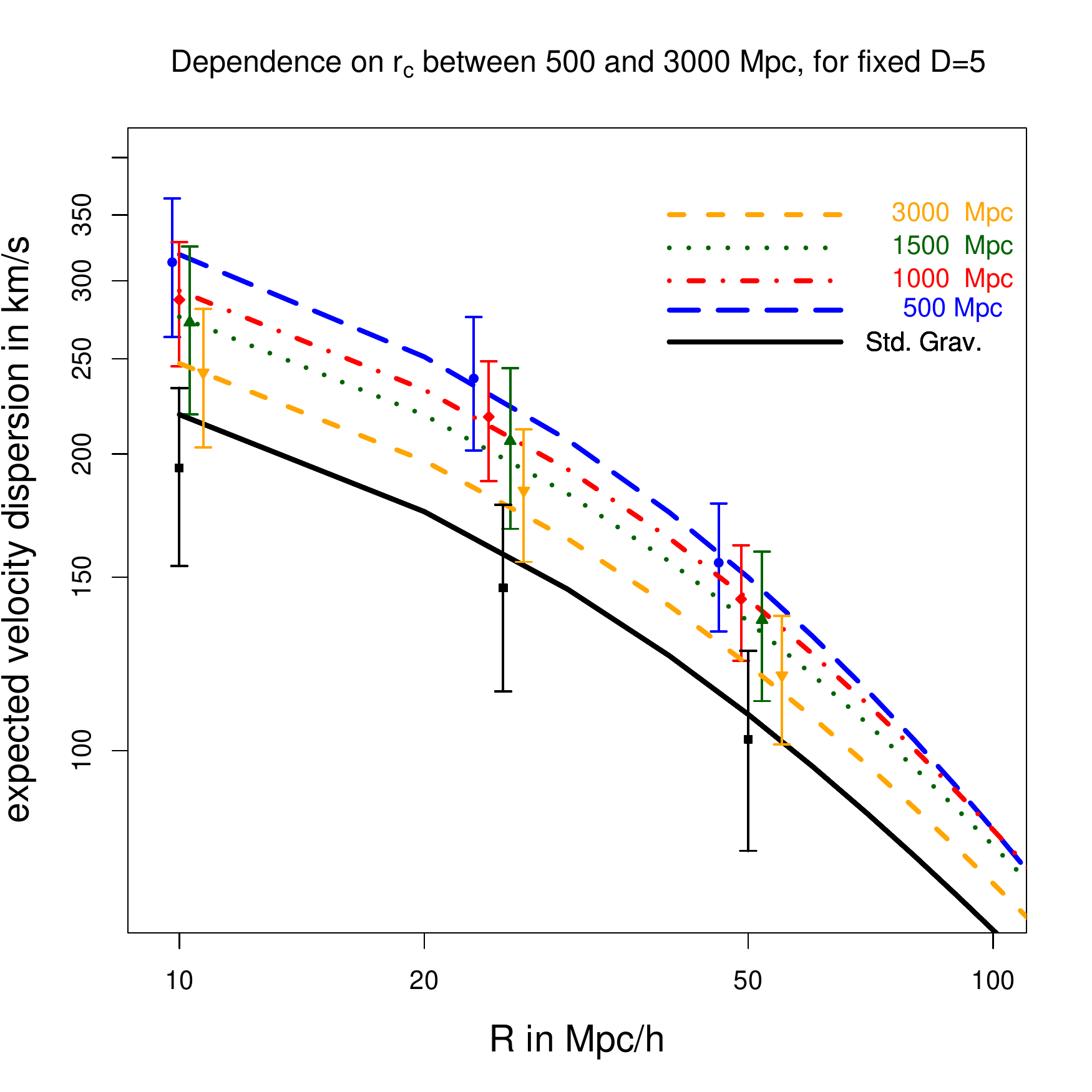} 
   \caption{The expected 1-dimensional peculiar velocity as a function of scale, using the Gaussian window function~(\ref{gauss}), for $D=5$ and, from top to bottom, 
   $r_c = 500$, 1000, 1500 and 3000~Mpc respectively in long-dashed (blue), dash-dotted (red), dotted (green), and short dashed (orange) curves (linear theory), and 
   circle, diamond, upright triangle, and inverted triangle points (simulations, with bootstrap-estimated error bars).  The standard gravity results are plotted as a solid (black) curve and square points.
   The points at a given scale are slightly offset from one another for readability. For this comparison, $\sigma_8[\Lambda\mbox{CDM}]=0.85$.}
   \label{fiveD}
\end{figure}

\section{Numerical simulations}
\label{numerics}

While the linear theory analysis should be applicable to calculate bulk flows on scales $\gsim 50h^{-1}$~Mpc, in this Section we
explicitly check this against N-body simulations to include non-linear effects. Note that for the sole purpose of this comparison, we use a slightly higher
primordial normalization for which $\sigma_8[\Lambda\mbox{CDM}]=0.85$. 

The N-body results presented here are obtained
by integrating the $\chi$ equation of motion~(\ref{pieqn}) exactly, using a particle mesh approach. This is a notable improvement over
earlier simulations presented in~\cite{markw}, where a spherical approximation was employed. Although the exact evolution agrees with the approximate results
reported in~\cite{markw} to within $\lsim 5$\% for the power spectrum over the relevant scales, we nonetheless
use the full code in the present work; a detailed comparison of the two approaches will be discussed elsewhere~\cite{marcos}.

We performed a series of particle-mesh simulations of $400h^{-1}$~Mpc boxes
on a $512^3$ grid with $512^3$ particles. To determine bulk flow statistics,  we perform a real-space average
of velocities over a large number of spheres centered on points randomly placed throughout 
our simulation output, randomizing and rerunning each $400h^{-1}$~Mpc box 2 or 3 times for each choice of parameters to gather
better statistics. However, it is worth noting that each box gives highly consistent results with each of the other boxes and with the ensemble.
We also use identical initial conditions across different parameter
choices to isolate the effects of the new gravitational physics from random fluctuations. 

For each sphere, we calculate a Gaussian-weighted average velocity,
\be
\langle v(R)[\mbox{one sphere}] \rangle = \frac{1}{N} \sum_i v_i \; \exp\left[ -(r_i/R)^2\right]\,,
\ee
where $r_i$ is the distance from the $i$th point from the randomly-selected
center, and $N = \sum_i  \; \exp\left[ -(r_i/R)^2 \right]$ is a normalization factor. This gives 3 independent 1-dimensional velocities for each sphere. All together, we average over $1200-4500$ different velocities for each parameter pair. We then calculate the standard deviation of all the sphere-averaged 1-dimensional
velocities to obtain the rms velocity.

Figures~\ref{peclin1000} and~\ref{fiveD} compare linear theory integration (lines) with N-body results (points).
Figure~\ref{peclin1000} explores the sensitivity to the number of extra dimensions, keeping $r_c$ fixed
at 1000~Mpc. (For this value of $r_c$, only the $D=5$ case satisfies the observational constraints discussed
in Sec.~\ref{ampcons}; these plots are made only to illustrate the sensitivity to $D$.) Figure.~\ref{fiveD} shows
the dependence on $r_c$, fixing the number of dimensions at $D=5$. In each case, we performed the calculations for $R= 10,25,$ and $50 h^{-1}$ Mpc. We generated the error bars by 
bootstrap resampling subsets of the measured velocities, computing the rms for each subset, and using this as a dataset for constructing an error estimate. 

The upshot is that these figures show excellent agreement between linear theory and
N-body simulations over the scales of interest. For bulk flows, the effects of non-linearities are basically absent, as hoped. This 
is in contrast with other velocity-related phenomena studied in~\cite{Percival:2008sh},
where the effects of non-linearities persist to scales $\sim 50 h^{-1}$ Mpc.

\section{Comparison with Data}
\label{data}

\subsection{Bulk Flows}

\begin{figure}[htbp] 
   \centering
   \includegraphics[width=0.5 \textwidth]{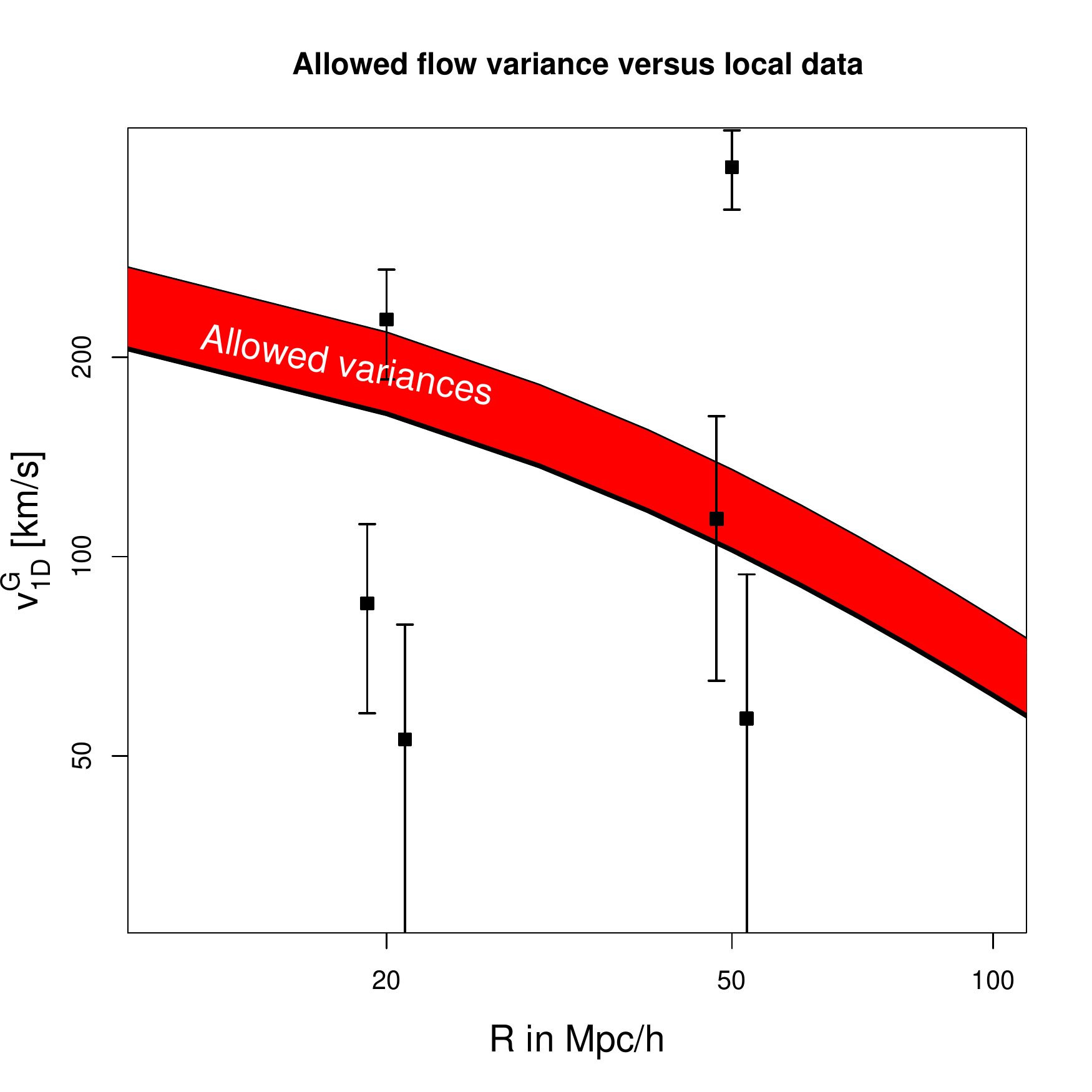} 
   \caption{Comparison of the data from~\cite{hudson1} (points) with the range of bulk flows
   achievable in Cascading Gravity models (shaded region, using Gaussian window function~(\ref{gauss})).
    The heavy curve defining the  bottom of the shaded region is the result for
   standard gravity for our fiducial cosmology. Note that the $20h^{-1}$ and $50h^{-1}$~Mpc data points are not independent.
   Using only the $50h^{-1}$ Mpc data points and comparing their
   fit to standard gravity to the fit to the maximum allowed velocity (which is reached
   for $D=10$, $r_c=2750$ Mpc), we find a $\Delta \chi^2 \simeq -4.6$.}
   \label{vdata}
\end{figure}

Figure~\ref{vdata} compares the range of expected 1-dimensional
velocity variances with the three local flow components
reported in~\cite{hudson1}. Our task in this Section is to quantify
the extent to which these data are more likely in Cascading Gravity as compared to GR.
We focus on the largest scales ($50h^{-1}$~Mpc) to minimize the influence
of non-linear structures. Moreover, the data on different scales are correlated and would require a careful treatment of covariances.
Note that all cosmological parameters are kept fixed in this analysis. We leave a comprehensive parameter likelihood analysis to future study.

\begin{figure}[htbp] 
   \centering
   \includegraphics[width=0.5 \textwidth]{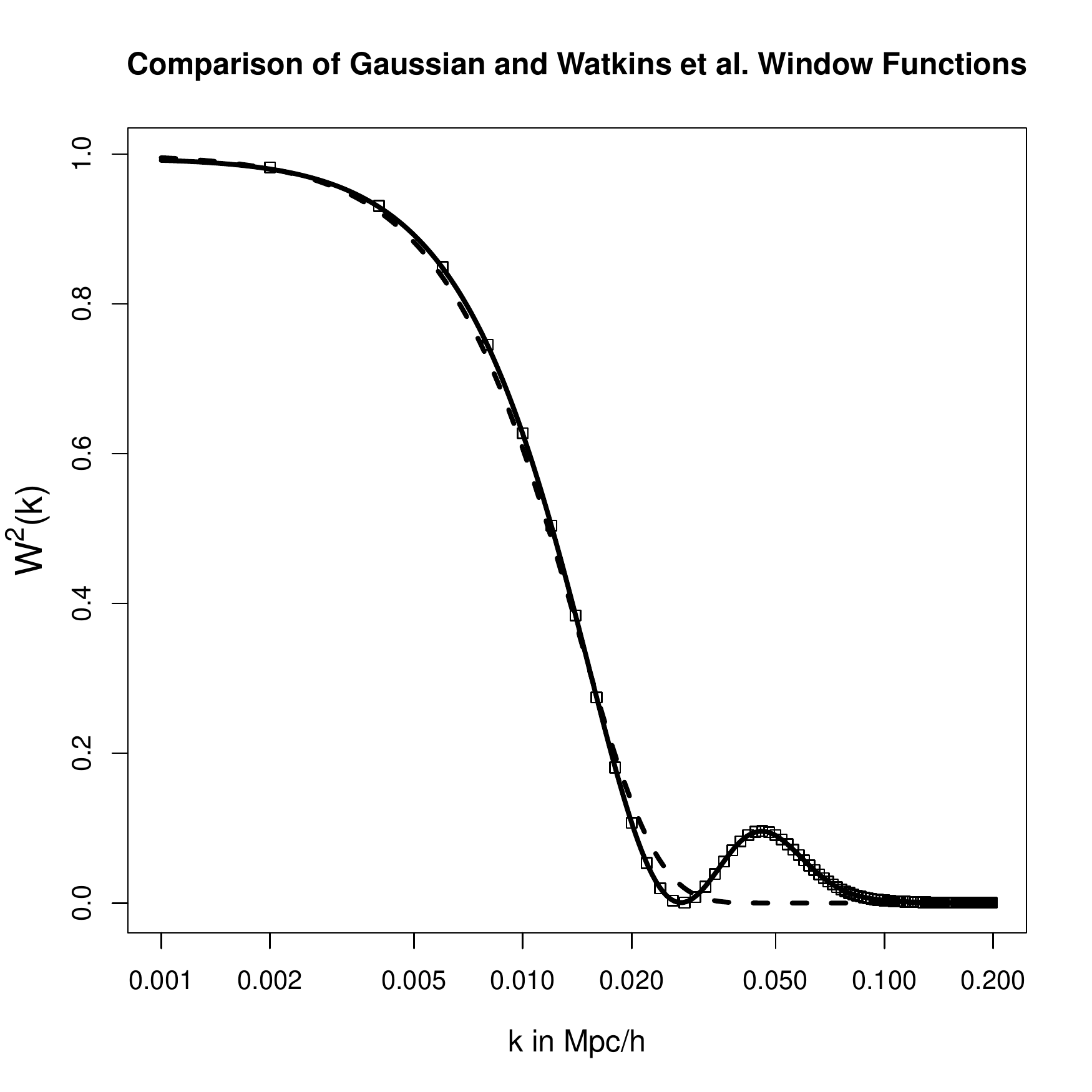} 
   \caption{Comparison of the Watkins {\it et al.} window function from~\cite{hudson1} (squares), 
   the fitting function given in~(\ref{winfit}) (solid line) and the Gaussian window function (dashed line) with $R=50 h^{-1}$~Mpc,
      normalized such that their value is unity for $k \rightarrow 0$.
   The difference between the results for the two different window functions is very small, of order $\sim 1$\%.}
   \label{wind}
\end{figure}

As mentioned earlier, for the comparison with data on $50h^{-1}$~Mpc scales we use the same window function as determined in~\cite{hudson1}
to analyze peculiar velocity surveys. This window function, shown in Fig.~\ref{wind}, is well-fitted by
\begin{widetext}
\be
\label{winfit}
W^{2,{\rm W}}(k) \simeq  3 \cdot \frac{1.86\cos(93.5 \, k)+0.0004\cosh(114.1\,k)-11.58\sin(2 \,k)+1.35\sinh(46\,k)}{4.89\cos(0.0012\,k)+0.73\,\cosh(121.3\,k)+1.45\sinh(66.6\,k)} \,.
\ee
\end{widetext}
Figure~\ref{wind} compares this fitting function with the actual window function of~\cite{hudson1} and
with our Gaussian window function. (In Fig.~\ref{vdata}, we instead use the Gaussian window function,
since~(\ref{winfit}) only applies on $50h^{-1}$~Mpc scales.)

The local bulk flow has three velocity components $v_i$, $i\in(1,2,3)$, measured by~\cite{hudson1}. Their
measurements have observational uncertainties $\sigma_i$, which
we assume represent independent Gaussian error bars. Although these components are in truth related by a covariance matrix,
for simplicity we assume that the data points are independent. This assumption has
only a small effect on the statistics \cite{mikeprivate}. We  compare these measurements
with the expected distribution of 1-dimensional velocities, which has vanishing mean and variance
$v_{1-{\rm dim}}^{\rm W}$. Assuming a Gaussian
probability distribution, we can compare the different likelihoods through
a simple $\chi^2$ statistic:
\be
\chi^2 \simeq  \sum_{i=1}^3  \frac{v_i^2}{\langle v_{1-{\rm dim}}^2 \rangle + \sigma_i^2} \,.
\ee
Recall that for our fiducial cosmology the standard gravity peculiar velocity variance is $v_{1-{\rm dim}}^{\rm W}[\rm{std\;grav}]  = 104$~km/s.
We wish to compare this with the maximum allowed variance in our cascading gravity models, $v_{1-{\rm dim}}^{\rm W}[\rm{mod\;grav}]  = 137$~km/s,
achieved with $D=10$ and $r_c = 2750$~Mpc. Comparing the $\chi^2$ values, we obtain
\be
 \Delta \chi^2_{\rm{eff}} \simeq -4.6
\ee
in favor of the Cascading Gravity model, or $-1.54$ per degree of freedom.

Although appreciable, this improvement in the fit does not by itself warrant
the inclusion of two new parameters ($r_c$ and $D$) in a strict sense. However such parameter counting is often
misleading. The underlying goal of long-distance modification of gravity is to relate these parameters to an existing one,
$\Lambda$. In a self-consistent analysis of cosmological predictions in Cascading Gravity, it is conceivable that
these parameters are in fact related to one another. For the purpose of this work, we can state conservatively that
current peculiar velocity data tantalizingly hint at gravitational physics beyond
$\Lambda$CDM but do not require it.

Meanwhile, on the very large, $300 h^{-1}$ Mpc scales relevant to the result of~\cite{kash1}, our modifications to gravity cannot account 
for the observed bulk flow. For our  $D=10$ and $r_c = 2750$~Mpc model, we find 
$v_{1-{\rm dim}}^{\rm G}[{\rm mod\;grav}] = 29$~km/s, versus 24~km/s for
$\Lambda$CDM. 


\subsection{Bullet Cluster}

As mentioned in the Introduction, recent hydronamical simulations of the Bullet Cluster have shown that an initial velocity
of  $v\simeq 3000$ km/s is required  when the cluster and subcluster are $\sim5$ Mpc apart to best reproduce
X-ray observations. Lee and Komatsu~\cite{Lee:2010hj} have recently estimated that the probability of
such an initial velocity in the standard $\Lambda$CDM framework is between $3.3\times10^{-11}$ and
$3.6\times10^{-9}$ --- respectively 6.5 and 5.8$\sigma$ from the mean of cluster velocities ---,
 where the uncertainty comes from the evolution of velocities with redshift.

Most of the infall velocity is caused by the gravitational attraction of the main cluster,
which is estimated to have a mass of $10^{15} M_{\odot}$ (That this is a good approximation for such massive clusters
is established in~\cite{Lee:2010hj}). We therefore compute the infall velocity in our models by treating the clusters as point particles
released from rest from some large initial separation $\sim30$ Mpc and integrate their dynamics down to a separation
of 5~Mpc as in~\cite{Lee:2010hj}. For simplicity, we assume head-on collision. The gravitational potential due to the large cluster is
obtained by solving~(\ref{pieqn2}) for $\chi$ and the standard Poisson equation for $\Psi_{\rm N}$ for a point particle of
mass $10^{15} M_{\odot}$. The motion of the subcluster is then obtained from the full gravitational potential given by~(\ref{dyn}).
We also include in quadrature the effect of the enhanced bulk flow on 5~Mpc scales; however, the final velocity is almost entirely determined
by the infall calculation.

We should expect the enhancement to scale as $G_{\rm eff}^{1/2}$, where $G_{\rm eff}$ is the effective Newton's constant
enhanced by the additional scalar forces in our models --- see~(\ref{deltaevolve}). In the last stages of infall, however, 
the enhancement is somewhat tamed by the Vainshtein mechanism. A related point is that the enhanced force in modified gravity
does not substantially reduce the velocity at 5~Mpc necessary to explain the merger velocity. Because of Vainshtein screening, the final
velocity of a particle falling from 5~Mpc to the center of a $10^{15} M_{\odot}$ cluster differs by at most 1\% compared with the infall
in standard gravity.

We find that initial velocities are 14\% to 27\% larger in our framework, 
with the smallest difference coming from $D=5$ and the largest for $D=10$,
again assuming the minimal value of $r_c$ allowed by constraints on $\sigma_8$:
$r_c= 1665$ and $2750$~Mpc, respectively.
These enhancements shift the mean of the probability density function for initial velocities calculated in~\cite{Lee:2010hj},
which is a function of $\log v$. The larger bulk flow component at 5~Mpc also slightly widens the variance of the distribution.

The end result is that the required initial velocities for the Bullet 
Cluster merger are much more likely in our model. Focusing on the $D=10$, $r_c= 2750$~Mpc case,
the probability of an initial velocity of $3000$ km/s at $z=0$ is increased to $6.6\times10^{-7}$ ---
2.0$\times 10^4$ times more likely than the $\Lambda$CDM value; in other words, a change from
a 6.5$\sigma$ to a 4.8$\sigma$ event. The probability of that
initial velocity at $z=0.5$, closer to the actual merger redshift of $z=0.296$, is increased to
$5.1\times10^{-5}$, again more than $10^4$ times as probable as the $\Lambda$CDM result --- 
a shift from 5.8$\sigma$ to 3.9$\sigma$.
If we follow~\cite{Lee:2010hj} and also consider the probability of finding an initial
velocity of 2000~km/s, the probability of such an occurrence in our model becomes 0.8\% at $z=0$ and
14.2\% at $z=0.5$, respectively 257 and 65 times more likely than the $\Lambda$CDM expectation 
 (that is, respectively 2.4 and 1$\sigma$ in modified gravity, versus 4 and 2.9$\sigma$ 
in standard gravity). This significant increase in probability is another tantalizing hint of new gravitational physics.

\section{Conclusions}

In this paper, we have explored how peculiar velocities are affected in a broad class of IR-modified
gravity theories called brane-induced gravity. On the scales of interest, these theories admit a local
$4D$ description in terms of weak-field gravity plus $D-4$ scalar fields coupled to the trace of the matter stress-tensor.
These scalar degrees of freedom effectively strengthen the gravitational attraction at late times and speed up 
structure formation. As a result, peculiar velocities are systematically larger than those
expected in standard gravity. Comparisons between N-body simulations and linear theory
calculations show that linear theory gives an excellent description of the physics of bulk flows.

We have found that large-scale bulk flows can be enhanced up to $\sim 40\%$ relative to $\Lambda$CDM cosmology.
The enhancement is limited by observational constraints on $\sigma_8$, the tightest limit coming from X-ray
cluster abundance. The predicted peculiar velocities alleviate the tension with recent observations of the bulk flow on $50h^{-1}$~Mpc scales
by Watkins {\it et al.}~\cite{hudson1,hudson2}, from a $\approx 3\sigma$ discrepancy in $\Lambda$CDM gravity to a $\approx 2\sigma$ difference
in Cascading Gravity. The agreement between theory and data is improved by $\Delta \chi^2 \simeq -4.6$ in our model. Although modest, this improvement 
offers further motivation for more accurate bulk flow observations.
Peculiar velocities are also enhanced on much larger scales ($\sim 300h^{-1}$~Mpc) probed by kinetic Sunyaev-Zel'dovich observations, but this is insufficient to explain the
enormous bulk flow inferred by~\cite{kash1,kash2}. 

Violent merging systems, such as the Bullet Cluster, are much more probable in Cascading Gravity. Drawing
on the recent analysis of~\cite{Lee:2010hj}, we have found that the occurrence of the Bullet Cluster in our theories is $\gtrsim 10^4$
times more likely than in standard gravity.

We are pursuing various parallel tracks to improve upon the preliminary analysis presented here.
As mentioned above, the tightest constraint on our models comes from cluster counts. Through N-body
simulations and various semi-analytical techniques, we are currently determining the halo mass function in Cascading Gravity as a
function of $D$ and $r_c$ \cite{marcos}. This will allow
a more accurate comparison with X-ray data. With regards to the Bullet Cluster, we can use similar simulations
to calculate the distribution of initial velocities for subclusters, following the standard gravity analysis
of~\cite{Lee:2010hj}.

On the theoretical side, the calculations presented here rely on some phenomenological input. While we are confident that our results
capture the essence of Cascading cosmology predictions, they do not derive from a rigorous cosmological analysis of the complete
higher-dimensional theory. An important first step in this direction would be to obtain the decoupling limit of Cascading Gravity while keeping the
non-linearities in all scalar degrees of freedom. This would allow a derivation of the modified Friedmann equation, as in~\cite{galnic} for
DGP, as well as the perturbation equations in the Newtonian regime. This is work in progress~\cite{kurtmarkdan}.

The observations of large bulk flows and violent cluster mergers offer tantalizing evidence that structure
is evolving more rapidly than predicted by $\Lambda$CDM cosmology. As we extend our measurements
of large-scale bulk flows beyond our local region and discover an increasing number of merging systems, it will become clear 
whether these are statistical flukes or the first indication of a new realm of gravitational physics on cosmological distances
in the late universe.

\vspace{0.5in}
\section*{Acknowledgements}
\vspace{-0.2in}
The authors would like to thank N.~Afshordi, M.~Hudson, L.~Hui, B.~Jain, E.~ Komatsu, M.~Lima, R.~Sheth, C.~Springob, M.~Trodden, W.~Percival and R.~Watkins for helpful discussions, F.~Schmidt for helpful comments and coding advice, and especially H.~Feldman for providing his window function and for invaluable help with peculiar velocity calculations. The work of M.W. was supported by the Perimeter Institute for Theoretical Physics.  Research at Perimeter Institute is supported by the Government of Canada through Industry Canada and by the Province of Ontario through the Ministry of Research \& Innovation. M.W. is grateful to the Center for Particle Cosmology for hospitality while part of this work was completed. The work of J.K. was supported by NSERC of Canada and funds from the University of Pennsylvania.


\begin{thebibliography}{99}
\bibitem{footnote1}
As a technical remark, the tensor structure in~(\ref{amp}) arises
when one adds higher-dimensional Einstein-Hilbert terms localized on the brane to excise the ghost mode discussed in the Introduction~\cite{cascade2}.
If one also adds a tension on the 3-brane, then at least in the $D=6$ case the exchange amplitude around this solution becomes tension-dependent~\cite{cascade1}.
\end{thebibliography}
\end{document}